\documentclass{ws-ijmpb}

\usepackage{graphicx}
\usepackage{amsmath}
\usepackage{amsfonts}
\usepackage{amssymb}
\usepackage{mathptmx}      
\usepackage{latexsym}

\begin{document}

\markboth{Myong Chol Pak, Kwang-Il Kim, Hak Chol Pak, Chol Won Ri, Sang Jun Cha}
{Spin Transport in the heterogeneous structure containing Topological Insulator and Diluted Magnetic Semiconductor}

%
\catchline{}{}{}{}{}
%

\title{Spin Transport in the heterogeneous structure containing Topological Insulator and Diluted Magnetic Semiconductor}

\author{Myong Chol Pak\footnote{\email{myongcholpak@163.com}}, 
        Kwang-Il Kim,
        Hak Chol Pak,
        Chol Won Ri,  
    	Sang Jun Cha}

\address{
           Department of Physics, Kim Il Sung University, Ryongnam Dong, Taesong District \\
           Pyongyang, Democratic People's Republic of Korea}
       
\maketitle

\begin{history}
	\received{28 November 2020}
\end{history}

\begin{abstract}
	In this paper, we discuss spin transport in topological insulator (TI) and diluted magnetic semiconductor (DMS) heterogeneous structures. 
	
	In DMS / FM (Ferromagnetic Metal) heterogeneous structure, the spin injection efficiency changes according to the electric field and Fermi energy were investigated. The higher electric field, the stronger spin injection efficiency, and its velocity of increase gets lower and approaches to the equilibrium state. Additionally, the higher interface conductivity, the weaker spin injection efficiency, and the transmission due to Fermi energy of spin up and spin down is different from each other. In the same structure, the spin injection efficiency changes depending on the magnetic field, and the spin injection efficiency vibrates sensitively according to the magnetic field. 
	
	In TI / DMS heterogeneous structure, the spin current changes according to magnetic field were investigated. Here, when the magnetic field is low, the spin current oscillates, and as the magnetic field increases, the vibration is attenuated. It also decreases with increasing temperature and weakens vibrations. This is due to the competitive effect of the chiral properties of Dirac type quasiparticles in the topological insulator and the unique properties of exchange interaction between electrons and ions in DMS. This result allows us to expect the possibility of spintronic devices with high sensitivity to magnetic field.
\end{abstract}

\keywords{Topological Insulator; Diluted Magnetic Semiconductor; Spin Transport}

\section{Introduction} \label{sec:intro}
Recently, spintronics has developed rapidly, taking advantage of its own properties in science and technology \cite{bib1,bib2}. Particular attention is paid to the integration of magnetic and semiconductor materials that can add new functions to electronic devices, as well as giant magnetoresistance (GMR) effect for large-scale savings of data. Popular materials here are diluted magnetic semiconductor (DMS) and topological insulator (TI).

DMSs are semiconductors doped with magnetic impurities, which have randomly distributed, inactive electrical properties, and do not have any long-range spatial spin ordering \cite{bib3,bib4}. The essential property in this material is the short-range antiferromagnetic superexchange interaction between Mn and the heavier 3d transition metal ions or the long-range dipolar interaction between rare earth magnetic moments \cite{bib5}. Using DMS, the mechanism of bound magnetic polaron \cite{bib6} and qubit formation at single quantum dot \cite{bib7} have also been studied. Plus, efficient spin injection and spin transport characteristics have been studied by making the multilayer structure using a diluted ferromagnetic semiconductor, and the effects of magnetization manipulation by deformation, light, electric field, and spin current on metal spintronics have been studied\cite{bib8}. In addition, the study of optical spin injection characteristics from DMSs to quantum dots has been intensified \cite{bib9}.

In order to consider the effect of magnetic field on the critical super current in DMS heterogeneous structures, the research has been in progress widely \cite{bib10}. However, since the change range of the critical super current due to the influence of the magnetic field is limited, finding a new physical quantity that can be changed sensitively to the magnetic field is a problem. On the other hand, after the fact that spin can be controlled \cite{bib10,bib11}, research is actively conducted. The most important problem here is to form the spin polarization state and transport the polarized spin state. However, the problem of transporting spin-polarized states experimentally is still under study. Therefore, the theoretical study on the transport of spin polarization has been actively conducted\cite{bib12,bib13}.  Here, semiconductor spintronic devices have attracted much attention due to their long relaxation time and long spin transport length \cite{bib14,bib15}. In addition, in order to design and manufacture high performance spin devices, it is necessary to select heterogeneous structures with excellent properties in the spin transport.

Meanwhile, a topological Insulator (TI) is a three-dimensional insulator with a strong spin-orbital coupling. In other words, the TI is a material that exhibits an insulator state when viewed in bulk, but a conductive state when viewed in surface \cite{bib16,bib17}. This state is caused by time inversion symmetry. Currently, research projects are being actively conducted to apply topological insulators to reality. Major advances have been made in the field of quantum information because the presence of Majorana Fermion is found at the interface between topological insulators and superconductors \cite{bib18}. In addition, we have attracted much attention in terms of anomalous magnetoresistance effect in topological insulator heterogeneous structures \cite{bib19}. Realizing magnetic switch based on spin structure and momentum distribution of electrons passing through topological insulator p-n junction (TIPNJ), many research projects are also underway \cite{bib20,bib21}. Also, nonlinear spin currents are generated in the three-dimensional TI with Rashba / Dresselhaus interaction \cite{bib22,bib23}, and it is also possible to switch between spin current and charge current in the surface state of topological insulator \cite{bib24,bib25}. Additionally, equilibrium and non-equilibrium transport studies through two-dimensional Josephson junctions of TIs \cite{bib26}, thermoelectric transport studies \cite{bib27}, research projects to realize spin structures with chiral properties \cite{bib28} and research into the unique phenomena in TI thin films has been actively conducted \cite{bib29}. And, various unusual effects have been proposed, such as the quantized anomalous Hall effect \cite{bib30}, the Aharanov-Bohm effect generated in the surface state of the TI nanoribbons \cite{bib31}, and the inverse Edelstein effect that occurs during spin injection in the topological kondo insulator\cite{bib32}. However, studies on the spin transport in heterogeneous structure between DMSs and TIs have not been discussed.

Since DMS is the semiconductor alloy in which magnetic atoms are substituted, their transport properties, such as the tunneling magnetoresistence and spin polarization, change with the external magnetic field. Therefore, we analyze the spin injection efficiency in the heterogeneous structure made of DMS and ferromagnetic metal (FM) by the spin injection efficiency formula in the heterogeneous structure composed of ferromagnetic metal and semiconductor, and consider the spin injection efficiency change according to the magnetic field. Also, we attempt to elucidate the factors affecting spin transport in heterogeneous structures, including TI and DMS.

This paper is organized as follows. Section 2 gives a methodology for calculating spin injection efficiency in heterogeneous structures including semiconductors, and discusses interfacial conductivity and spin injection efficiency in heterogeneous structure made of DMS and FM. Then, the spin injection efficiency as a function of the magnetic field is also examined how the magnetic field changes. Section 3 discusses spin transport properties in TI/DMS heterogeneous structure and investigates their temperature and magnetic field dependence. In Section 4, summary and conclusion are presented.

\section{Spin Injection in DMS / FM Heterogeneous structure} \label{sec:1}

Before investigating DMS and FM heterogeneous structure, we first briefly mention the methodology for calculating spin injection efficiency in the semiconductor and ferromagnetic metal heterogeneous structure \cite{bib11}.

Generally, if we assume that there is no space charge and the material is homogeneous, the currents for the up (down)-spin are as follows.
\begin{equation}
\label{eq:sc1}
\begin{split}
& \vec j_{ \uparrow ( \downarrow )}  = \vec j_{drift}  + \vec j_{diff}   \\
& \vec j_{drift}  = \sigma _{ \uparrow ( \downarrow )} \vec E  \\
& \vec j_{diff}  = eD_{ \uparrow ( \downarrow )} \nabla n_{ \uparrow ( \downarrow )}   \\
& \sigma _{ \uparrow ( \downarrow )}  = \sigma _{ \uparrow ( \downarrow )}^0  + n_{ \uparrow ( \downarrow )} ev_{ 
	\uparrow ( \downarrow )}  \\
\end{split}
\end{equation}
where $D_{ \uparrow ( \downarrow )}$ is the up(down)-spin electron diffusion constant , $\sigma _{ \uparrow ( \downarrow )}$ is up(down)-spin conductivity and   $\sigma _{ \uparrow ( \downarrow )}^0$ , $v_{ \uparrow ( \downarrow )}$ are the conductivity in the absence of spin polarization, the mobility, respectively.
Also, the continuity equations for up (down)- spin electrons can be as 
\begin{equation}
\label{eq:sc2}
{{\partial n_{ \uparrow ( \downarrow )} } \over {\partial t}} =  - {{n_{ \uparrow ( \downarrow )} } \over {\tau _{ \uparrow  \downarrow ( \downarrow  \uparrow )} }} + {{n_{ \downarrow ( \uparrow )} } \over {\tau _{ \downarrow  \uparrow ( \uparrow  \downarrow )} }} + {1 \over e}\nabla  \cdot \vec j_{ \uparrow ( \downarrow )} 
\end{equation}

For a homogeneous without space charge $n_{ \uparrow ( \downarrow )}$ satisfy the condition $n_ \uparrow   + n_ \downarrow   = 0$ , so that we can obtain the following equation.

\begin{equation}
\label{eq:sc3}
\begin{split}
& \nabla ^2 (n_ \uparrow   - n_ \downarrow  ) + {v \over {eD}}eE\nabla (n_ \uparrow   - n_ \downarrow  ) - {{n_ \uparrow   - n_ \downarrow  } \over {L^2 }} = 0  \\ 
& v = {{\sigma _ \uparrow  v_ \downarrow   + \sigma _ \downarrow  v_ \uparrow  } \over {\sigma _ \uparrow   + \sigma _ \downarrow  }}  \\
& D = {{\sigma _ \uparrow  D_ \downarrow   + \sigma _ \downarrow  D_ \uparrow  } \over {\sigma _ \uparrow   + \sigma _ \downarrow  }}  \\
& L = \sqrt {D\tau _s }  \\
\end{split}
\end{equation}

And the relationship between electrochemical potential and nonequilibrium carrier density in a highly degenerate system is 

\begin{equation}
\label{eq:sc4}
n_{ \uparrow ( \downarrow )}  = eN_{ \uparrow ( \downarrow )} (\varepsilon _F )(\mu _{ \uparrow ( \downarrow )}  + e\psi )
\end{equation}
where $N_{ \uparrow ( \downarrow )} (\varepsilon _F )$ is the density of state of up(down)-spin at the Fermi energy and $\psi$ is related to the electric field, ${\bf{E}} =  - \nabla \psi $.
If we take into account Eq. (3), (4), the drift diffusion equation can be as
\begin{equation}
\label{eq:sc5}
\nabla ^2 \left( {\begin{array}{*{20}c}
	{\mu _ \uparrow  }  \\
	{\mu _ \downarrow  }  \\
	
	\end{array} } \right) = \frac{1}
{{L^2 }}\left( {\begin{array}{*{20}c}
	{\frac{{\sigma _ \downarrow  }}
		{{\sigma _ \uparrow   + \sigma _ \downarrow  }}} & {\frac{{ - \sigma _ \downarrow  }}
		{{\sigma _ \uparrow   + \sigma _ \downarrow  }}}  \\
	{\frac{{ - \sigma _ \uparrow  }}
		{{\sigma _ \uparrow   + \sigma _ \downarrow  }}} & {\frac{{\sigma _ \uparrow  }}
		{{\sigma _ \uparrow   + \sigma _ \downarrow  }}}  \\
	
	\end{array} } \right)\left( {\begin{array}{*{20}c}
	{\mu _ \uparrow  }  \\
	{\mu _ \downarrow  }  \\
	
	\end{array} } \right)
\end{equation}
If we assume that there are FM in the range of $x < 0$  and semiconductor in the range of $x > 0$ , the solution of Eq.(5) in FM is

\begin{equation}
\label{eq:sc6}
\frac{1}
{{eJ}}\left( {\begin{array}{*{20}c}
	{\mu _ \uparrow  }  \\
	{\mu _ \downarrow  }  \\
	
	\end{array} } \right) = \frac{x}
{{\sigma _ \uparrow ^f  + \sigma _ \downarrow ^f }}\left( {\begin{array}{*{20}c}
	1  \\
	1  \\
	
	\end{array} } \right) + c\left( {\begin{array}{*{20}c}
	{\frac{1}
		{{\sigma _ \uparrow ^f }}}  \\
	{\frac{{ - 1}}
		{{\sigma _ \downarrow ^f }}}  \\
	
	\end{array} } \right)\exp (x/L^{(f)} )
\end{equation}
Additionally, the up (down)-spin charge densities in semiconductor must be satisfied the local charge neutrality condition and Eq. (5), so that the carrier density $n_{ \uparrow ( \downarrow )}$  can be as 
\begin{equation}
\label{eq:sc7}
n_{ \uparrow ( \downarrow )}  =  + ( - )A\exp (\frac{{ - x}}
{{L_d }})
\end{equation}
Therefore, from Eq. (6) the electrochemical potential is
\begin{equation}
\label{eq:sc8}
\mu _{ \uparrow ( \downarrow )}  = k_B T\ln [1 + ( - )\frac{{2A\exp ( - x/L_d )}}
{{n_0 }}] + eEx - B
\end{equation}

According to the condition that the current for all spin direction is continuous across the interface and such current is related to the change of spin dependent electrochemical potential, the following boundary conditions take place.

\begin{equation}
\label{eq:sc9}
\begin{gathered}
j_ \uparrow  (0^ -  ) = G_ \uparrow  [\mu _ \uparrow  (0^ +  ) - \mu _ \uparrow  (0^ -  )] \hfill \\
j_ \downarrow  (0^ -  ) = G_ \downarrow  [\mu _ \downarrow  (0^ +  ) - \mu _ \downarrow  (0^ -  )] \hfill \\
j_ \uparrow  (0^ -  ) - j_ \downarrow  (0^ -  ) = j_ \uparrow  (0^ +  ) - j_ \downarrow  (0^ +  ) \hfill \\ 
\end{gathered} 
\end{equation}
If the spin polarization is not very large in the semiconductor, we obtain
\begin{equation}
\label{eq:sc10}
\begin{gathered}
\alpha (x) = \left[ {{{L^{(f)} } \over {(1 - p_f^2 )\sigma _f }} + {{L^{(s)} } \over {\sigma _s }} + {{G_ \uparrow   + G_ \downarrow  } \over {4G_ \uparrow  G_ \downarrow  }}} \right]^{ - 1} \\
\;\;\;\;\;\;\;\;\;\; \times \left[ {{{p_f L^{(f)} } \over {(1 - p_f^2 )\sigma _f }} + {{G_ \uparrow   - G_ \downarrow  } \over {4G_ \uparrow  G_ \downarrow  }}} \right]e^{ - x/L^{(s)} }
\end{gathered}
\end{equation}
where $ \sigma _f  = \sigma _ \uparrow ^f  + \sigma _ \downarrow ^f $ is the conductivity of FM, $p_f  = (\sigma _ \uparrow ^f  - \sigma _ \downarrow ^f )/(\sigma _ \uparrow ^f  + \sigma _ \downarrow ^f )$ is the spin polarization of FM, and $L^{(f)}$ is the spin diffusion length of FM. Also, $L^{(s)}$ and $\sigma_s$ are the spin diffusion length and conductivity of the semiconductor. In addition, $G_{ \uparrow ( \downarrow )}$ is interfacial conductivity of up (down)-spin and it is directly related to Hamiltonian of DMS through Landauer's formula \cite{bib37}. From these results, it can be seen that if the interfacial conductivity is accurately calculated in heterogeneous structures containing DMS, the spin injection efficiency of FM and DMS heterogeneous structures can be considered. At the same time,  $L^{(s)}$ and $\sigma_s$ are changed to the spin diffusion length and conductivity of DMS,
\begin{equation}
\label{eq:sc11}
\begin{gathered}
\alpha (x) = \left[ {{{L^{(f)} } \over {(1 - p_f^2 )\sigma _f }} + {{L^{(DMS)} } \over {\sigma _{DMS}}} + {{G_ \uparrow   + G_ \downarrow  } \over {4G_ \uparrow  G_ \downarrow  }}} \right]^{ - 1} \\
\;\;\;\;\;\;\;\;\;\; \times \left[ {{{p_f L^{(f)} } \over {(1 - p_f^2 )\sigma _f }} + {{G_ \uparrow   - G_ \downarrow  } \over {4G_ \uparrow  G_ \downarrow  }}} \right]e^{ - x/L^{(DMS)} }
\end{gathered}
\end{equation}

We now consider spin injection efficiency by setting ${\rm{Cd}}_{{\rm{1}} - {\rm{x}}} {\rm{Mn}}_{\rm{x}} {\rm{Te}}$ as DMS and obtaining interfacial conductivity in DMS / FM heterogeneous structure.

If the external magnetic and electric fields act along z-axis, the effective Hamiltonian is
\begin{equation}
\label{eq:sc12}
\begin{gathered}
H\psi  = E\psi \\
H = {{({\bf{p}} + e{\bf{A}})^2 } \over {2m^* }} + g_e \mu _B B - eEz + V_{ex} \\
{\bf{A}} = {\bf{B}} \times {\bf{r}}/2 = ( - y,x,0)B/2
\end{gathered}
\end{equation}
where $m^*$ is the effective mass of the electron, ${\bf{p}}$  is the momentum operator, ${\bf{r}} = \left( {{\rm{x,y,z}}} \right)$ is the position vector of the electron, $g_e$ is the Lande factor of the electron, ${\bf{E}} = (0,0,E)$ is the external electric field, and ${\bf{A}}$ is the vector potential. The exchange interaction term $V_{ex}$ is a exchange interaction between the electron and the magnetic ion ${\rm{Mn}}^{{\rm{2}} + }$,
\begin{equation}
\label{eq:sc13}
V_{ex}  = J_{s - d} \left\langle {S_z } \right\rangle \sigma _z 
\end{equation}
where $J_{s - d}  =  - N_0 \alpha x_{eff}$,$\left\langle {S_z } \right\rangle  = S_0 B_J [Sg\mu _B B/k_B (T + T_0 )]$ and $B_J (x)$ is Brillouin function. And $N_0$ is the number of positive ions per unit volume and $S = 5/2$ is the spin of localized electrons in the ${\rm{Mn}}^{{\rm{2}} + }$ ions. Also $x_{eff}$ is the effective concentration of ${\rm{Mn}}$ , $T_0$ is the reduced temperature considering the effect of antiferromagnetic ${\rm{Mn}}$-${\rm{Mn}}$ bond on a single ion, $\mu_B$ is the Bohr magneton and $g = 2$ is the Lande factor of the ${\rm{Mn}}^{{\rm{2}} + }$ion. $k_B$ is Boltzmann constant and $\sigma _z  =  \pm {\rm{1/2}}$ is electron spin.

According to Bloch theorem, the electronic states are expressed through the discrete inverse lattice vectors $(k_x  + nK_x ,k_y  + mK_y )$.
\begin{equation}
\label{eq:sc14}
\psi ({\mathbf{r}}) = \frac{1}
{{\sqrt {L_x L_y } }}\sum\limits_{n,m} {C_{nm} (z)\exp [i(k_x  + nK_x )x +}+ i(k_y  + mK_y )y]
\end{equation}
Inserting Eq.(14) into Eq.(12),the result is 
\begin{equation}
\label{eq:sc15}
\frac{{p_z^2 }}
{{2m^* }}C_{n'm'} (z) + V_{eff}^{n'm'} (z)C_{n'm'} (z) = EC_{n'm'} (z)
\end{equation}
where
$$
\begin{gathered}
V_{eff}^{n'm'} (z) = \{ \frac{{\hbar ^2 }}{{2m^* }}[(k_x  + nK_x )^2  + (k_y  + mK_y )^2 ] + \frac{{\omega _c^2 m^* L^2 }}{{96}} + g\mu _B B - eEz + V_{ex} \} +\hfill \\
+ \frac{1}{{L_x L_y }}\int {dxdy} V({\mathbf{r}})C_{nm} (z)\exp [i(k_x  + nK_x )x + i(k_y  + mK_y )y]C_{n'm'}^{ - 1} (z) + \hfill \\
+ \sum\limits_{n,m} {\{ (1 - \delta _{nn'} )\delta _{mm'} ( - 1)^{n - n'} [\frac{{\omega _c^2 m^* L_x^2 }}{{16\pi ^2 (n - n')^2 }} -} - \frac{{i\hbar \omega _c L_x (k_y  + mK_y )}}{{4\pi (n - n')}}]+ \hfill \\
+ (1 - \delta _{mm'} )\delta _{nn'} ( - 1)^{m - m'} [\frac{{\omega _c^2 m^* L_y^2 }}
{{16\pi ^2 (m - m')^2 }} +\frac{{i\hbar \omega _c L_y (k_x  + mK_x )}}{{4\pi (m - m')}}]\}  \hfill \\ 
\end{gathered} 
$$
Eq.(15) is solved by the difference expression using Noumerov method\cite{bib34} and the interface conductivity is calculated by Landauer’s formula . Then, inserting this interface conductivity into Eq.(11), the spin injection efficiency can be investigated quantitatively.

If Co is chosen as the ferromagnetic metal, its parameters are $p_f  = 0.5$, $L^{(f)}  = 60{\rm{nm}}$ and $\sigma _f  = {\rm{8}}00(\rm{\Omega}{\rm{cm}})^{ - 1}$ \cite{bib38}. In the case of DMS, the spin diffusion length is reduced due to the exchange scattering caused by the magnetic impurities, so it is reduced to $L^{(DMS)}  = 20n{\rm{m}}$ and becomes $\sigma _{DMS}  = 3(\rm{\Omega}{\rm{cm}})^{ - 1}$ \cite{bib35}. The relative transmission as a function of Fermi energy for up spin and down spin is shown in Fig. 1 when the parameters are  $m^*  = 0.096m_0 $ ,$x_{eff}  = 0.045$  ,$N_0 \alpha  = 0.22{\text{eV}}$  ,$N_0 \beta  =  - 0.88{\text{eV}}$  ,$S_0  = 1.32$  ,$T_0  = 3.1{\text{K}}$, and $g_e^*  =  - 0.7$ \cite{bib33}.

\begin{figure}[h]%
	\begin{center}
	\includegraphics*[width=8cm,height=6.5cm]{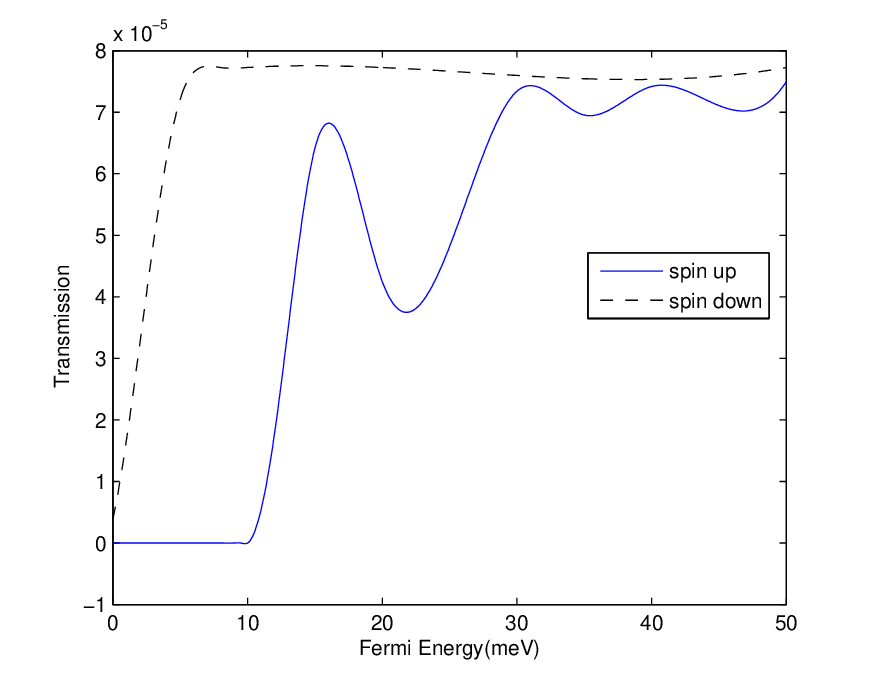}
	\caption{%
		Relative transmission as a function of Fermi energy.
	}
\end{center}
\end{figure}
As shown in Figure 1, in the FM / DMS heterogeneous structure, the relative transmission for the up spin and down spin are different from each other, which can be attributed to the exchange interaction term of the DMS \cite{bib10}.

Meanwhile, the change in spin injection efficiency as a function of the magnetic field is shown in Figure 2.

\begin{figure}[h]%
	\begin{center}
	\includegraphics*[width=8cm,height=6.5cm]{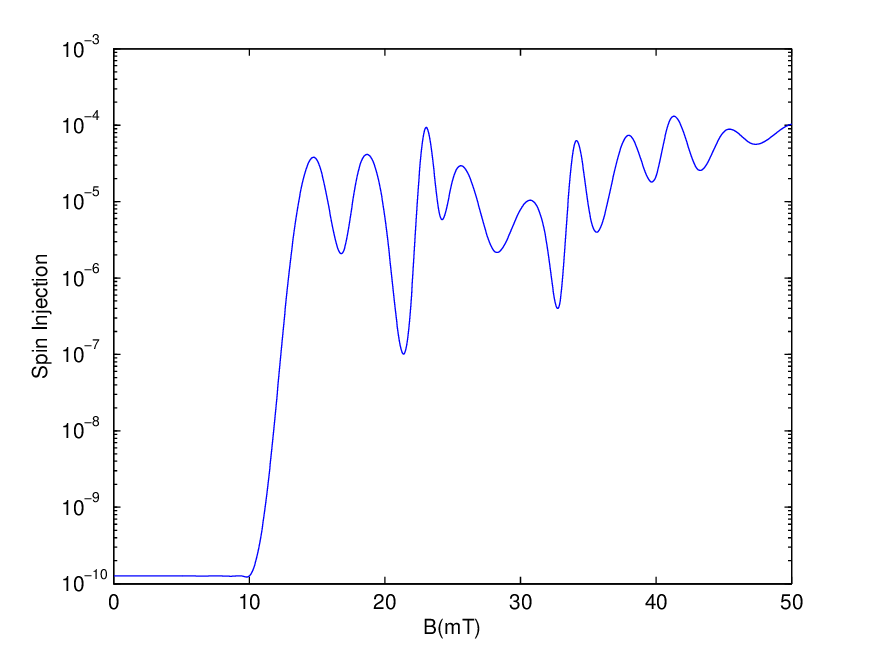}
	\caption{%
		Spin injection efficiency as a function of magnetic field.
	}

\end{center}
\end{figure}
Figure 2 shows that the spin injection efficiency vibrates rapidly with the magnetic field. This result is due to the effect of Landau levels and it enables us to expect the possibility of making the measuring device with keen sensitivity by using the spin injection efficiency in the same heterogeneous structure.

\section{Spin Injection in TI / DMS Heterogeneous structure}
\label{sec:2}
A Dirac cone on the surface of TI is represented by Hamiltonian\cite{bib36},
\begin{equation}
\label{eq:sc16}
H\left[ {v_F } \right] = \int {{{dk_x dk_y } \over {\left( {2\pi } \right)^2 }}\psi _{\bf{k}}^ +  (\hbar v_F \hat n \cdot \hat \sigma  \times {\bf{k}} + \mu _B \hat \sigma  \cdot {\bf{B}} - \mu \hat I)\psi _{\bf{k}} } 
\end{equation}
where $\hat \sigma$ and $\hat I$ are the Pauli matrix and the unit matrix, $\hat n$ is the unit vector perpendicular to the the surface of TI, $v_F$ is the Fermi velocity,   is Bohr magneton, $\mu$ is the chemical potential, and $\psi  = \left( {\psi _ \uparrow  ,\psi _ \downarrow  } \right)^T$ is the annihilation operator for Dirac spinor. Here $\uparrow \left(  \downarrow  \right)$ represents the direction of the quasiparticle spin polarized in the z-axis direction.

We now look at the spin current in TI/DMS heterogeneous structure shown in Figure 3. Zone I and II in Figure 3 should be discussed in terms of the chiral properties of Dirac-type quasiparticle, while Zone III is a DMS thin film zone and should be considered by traditional electrons subject to the Schrödinger equation. Therefore, in this heterogeneous boundary, the chiral properties of Dirac type quasiparticles and the properties of electrons in DMS are mixed, and thus new properties may be expected in the spin transport. 
\begin{figure}[h]%
\begin{center}
	\includegraphics*[width=8.3cm,height=6.5cm]{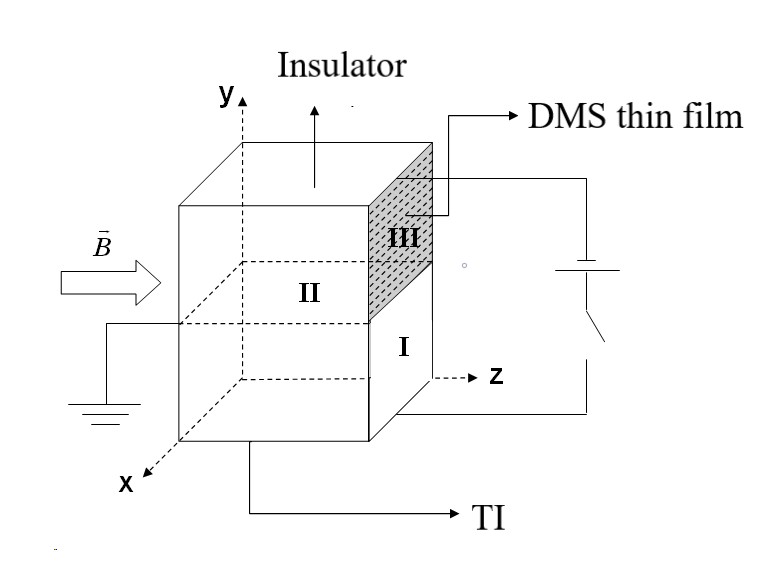}
	\caption{%
		Heterogeneous structure consisting of TI and DMS.
	}
\end{center}
\end{figure}

We discuss the effect of magnetic field on the spin current of TI and DMS heterogeneous structures by placing the DMS in zone III as ${\rm{Cd}}_{{\rm{1}} - {\rm{x}}} {\rm{Mn}}_{\rm{x}} {\rm{Te}}$, the previously discussed material. Using the Hamiltonian treated in Section 2 and considering the procedure in Ref.\cite{bib36}, the wave function in Zone III can be written as

\begin{equation}
\label{eq:sc17}
\psi _{{\rm{III}}}^{{\rm{DMS}}}  = [t_2 e^{ik_y^{\left( 1 \right)} y} ,t_3 e^{ik_y^{\left( 2 \right)} y} ]^T e^{ik_x x} /\sqrt 2 
\end{equation}
where
\begin{equation}
\label{eq:sc18}
\begin{gathered}
k_y^{\left( 1 \right)}  = \hbar ^{ - 1} [2m^* (\varepsilon  - J_{sp - d}  < S_z  >  - g_e^* \mu _B B) -\\
\,\,\,\,\,\,\,\,\,\,\,\, - (\hbar k_x  - eBy/2)^2 ]^{1/2}  - eBx/(2\hbar )
\end{gathered}
\end{equation}
\begin{equation}
\label{eq:sc19}
\begin{gathered}
k_y^{\left( 2 \right)}  = \hbar ^{ - 1} [2m^* (\varepsilon  + J_{sp - d}  < S_z  >  - g_e^* \mu _B B) -\\
\,\,\,\,\,\,\,\,\,\,\,\, - (\hbar k_x  - eBy/2)^2 ]^{1/2}  - eBx/(2\hbar )
\end{gathered}
\end{equation}
From this wave function, we can discuss spin current and junction properties. Given that there are no barriers to the junction, the boundary condition of the wave function is
\begin{equation}
\label{eq:sc20}
\psi _{{\rm{III}}}^{{\rm{DMS}}}  = c(\psi _{\rm{I}}  + \beta \psi _{{\rm{II}}} ),\;{\hbar  \over m}\partial _y \psi _{{\rm{III}}}^{{\rm{DMS}}}  = {{iv_1 \sigma _x } \over c}[\psi _{\rm{I}}  - \beta \psi _{{\rm{II}}} ]
\end{equation}
where $c$ is any real number, simply put $c=1$. If $c \to 0$ or $c \to \infty$, no heterogeneous structure is achieved. Then,$t_2$ and $t_3$ are represented as
\begin{equation}
\label{eq:sc21}
\begin{gathered}
t_2  =  - 4ie^{i\theta _k } \sin \left( {\theta _k } \right)\left( {u_k  + \alpha _1 v_k } \right)/D_2 \\
t_3  =  - 4ie^{i\theta _k } \sin \left( {\theta _k } \right)\left( {\alpha _2 u_k  + v_k } \right)/D_2
\end{gathered}
\end{equation}
where $\alpha _1  = \hbar k_y^{\left( 1 \right)} /\left( {mv_1 } \right)$, $\alpha _2  = \hbar k_y^{\left( 2 \right)} /\left( {mv_1 } \right)$ ,$D_2  = p_1  - ip_2 \exp \left( {i\theta _k } \right)$, $D_2  = p_1  - ip_2 \exp \left( {i\theta _k } \right)
$ and $p_2  = v_k \left( {1 + \alpha _1 \alpha _2 } \right) + 2u_k \alpha _2$. 

Now we can use $J_z  = (\hbar v_1 /2)\sum\nolimits_{k_x } {\left( {\alpha _1 \left| {t_2 } \right|^2  - \alpha _2 \left| {t_3 } \right|^2 } \right)}$ to evaluate the effect of the magnetic field on the spin current. By selecting the parameters of the DMS as described above, the spin current can be calculated from DMS thin film and TI structure (Figure 4).
\begin{figure}[h]%
	\begin{center}
	\includegraphics*[width=9cm,height=7cm]{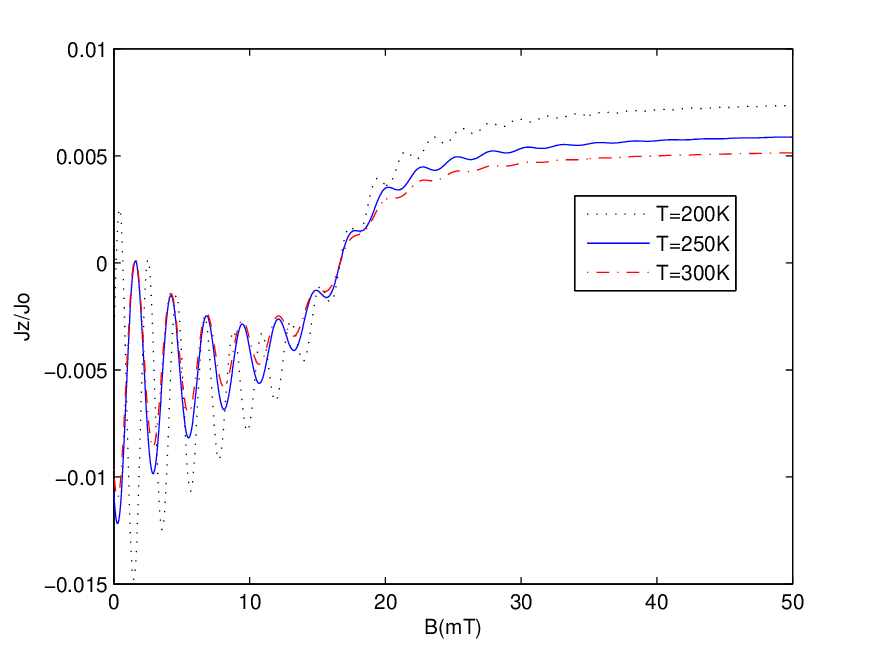}
	\caption{%
		Changes in spin current with magnetic field and temperature in TI / DMS Heterogeneous structure.
	}
	\end{center}
\end{figure}
As shown in Fig. 4, the spin current polarized in the z-axis shows a unusual behavior depending on the magnetic field. When the magnetic field is low, the vibration of the spin current is large, and as the magnetic field is increased, the vibration is attenuated.  This spin current also decreases with increasing temperature, and vibration also weakens. This is due to the competitive effect between the chiral property of Dirac-type quasiparticles in TI and $sp - d$  exchange interaction between electrons and ${\rm{Mn}}^{{\rm{2}} + }$ ions in DMS. When the magnetic field is applied, the vibration of spin current occurs in DMS due to Landau level. On the other hand, in TI, the spin current properties vary depending on the magnetic field due to the chiral properties of Dirac-type quasiparticles. As these two competitive effects occur at the same time, when the magnetic field grows, there is a peculiar behavior of mutual dependency. Also, as the temperature rises, the thermal movements become more active and will interfere with the directional spin current. Therefore, as shown in the figure, the spin current gradually weakens as the temperature increases.

\section{Summary}
\label{sec:3}
We have analyzed the spin injection efficiency properties according to Fermi energy in DMS / FM heterogeneous structures by spin injection efficiency formula and Landauer's formula, and the basic cause of the difference in relative transmission for up and down spins is found to be due to the exchange interaction term in DMS. By calculating the spin injection efficiency according to the magnetic field in the same structure, it was found that the measuring device with the most sensitive magnetic field sensitivity can be made. In addition, when magnetic field is applied to the heterogeneous structure composed of TI and DMS thin film, the spin effect is peculiar by the competition effect between the Landau level of DMS and the chiral characteristics of the Dirac-type quasiparticles in TI. This property will be of great significance in the future development and application of spintronics.

\section*{Acknowledgements}
It is a pleasure to thank Yong Hae Ko, Tae Hyok Kim and Kwang Il Kim for useful discussions. This work was supported by the National Program on Key Science Research of Democratic People's Republic of Korea (Grant No. 18-1-3).

\end{document}